%% file: LIZAI-XT.tex
\documentclass[3p,review]{elsarticle}

\usepackage{lineno,hyperref}
\usepackage{booktabs}
\usepackage{longtable}
\usepackage{geometry}
\usepackage{float}
\usepackage{pdflscape}
\usepackage{amsmath}
\usepackage{mathtools} 
\usepackage{cleveref}

\modulolinenumbers[5]

\journal{Journal of Artificial Intelligence (AIJ)}

\makeatletter
\def\ps@pprintTitle{%
 \let\@oddhead\@empty
 \let\@evenhead\@empty
 \let\@oddfoot\@empty
 \let\@evenfoot\@empty
}
\makeatother

\begin{document}

\begin{frontmatter}

\title{LizAI XT -- Artificial Intelligence-Powered Platform for Healthcare Data Management: A Study on Clinical Data Mega-Structure, Semantic Search, and Insights of Sixteen Diseases }


\author[lizai]{Trung Tin Nguyen\corref{mycorrespondingauthor}}
\author[tel]{Salomon M. Stemmer}
\author[lizai,harvard]{David R. Elmaleh\corref{mycorrespondingauthor}}

\address[lizai]{LizAI Inc., Newton, Massachusetts 02459, USA}
\address[tel]{Institute of Oncology, Davidoff Center, Rabin Medical Centre, Petah Tikva, Faculty of Medicine, Tel Aviv  University, Tel-Aviv, Israel}
\address[harvard]{Department of Radiology, Massachusetts General Hospital and Harvard Medical School, Boston, Massachusetts 02114, USA}

\cortext[mycorrespondingauthor]{Corresponding Authors: Trung Tin Nguyen: nguyen@lizai.co; David R. Elmaleh: delmaleh777@gmail.com. These authors contributed equally to this project.}

\begin{abstract}
AI-powered LizAI XT ensures real-time and accurate mega-structure of different clinical datasets and largely inaccessible and fragmented sources, into one comprehensive table or any designated forms, based on diseases, clinical variables, and/or other defined parameters. We evaluate the platform’s performance on a cluster of 4x NVIDIA A30 GPU 24GB, with 16 diseases – from deathly cancer and COPD, to conventional ones – ear infections, including a total 16,000 patients, $\sim115,000$ medical files, and $\sim800$ clinical variables. LizAI XT structures data from thousands of files into sets of variables for each disease in one file, achieving $>$95.0\% overall accuracy, while providing exceptional outputs in complicated cases of cancers (99.1\%), COPD (98.89\%), and asthma (98.12\%), without model-overfitting. Data retrieval is sub-second for a variable per patient with a minimal GPU power, which can significantly be improved on more powerful GPUs. LizAI XT uniquely enables fully client-controlled data, complying with strict data security and privacy regulations per region/nation. Our advances complement the existing EMR/EHR, AWS HealthLake, and Google Vertex AI platforms, for healthcare data management and AI development, with large-scalability and expansion at any levels of HMOs, clinics, pharma, and government. 
\end{abstract}


\end{frontmatter}


\section{Introduction}

Artificial intelligence (AI) aims to accelerate science and technology innovations, improve diagnostic accuracy, reduce physician burnout, enhance precision treatment, and improve management across levels from clinics, insurance companies, to government, in healthcare. Despite the recent hype and technical advancements in AI applications, some of which are invented by OpenAI~\cite{chatgpt}, Google~\cite{vertex_ai}, Amazon~\cite{aws_healthlake}, and Microsoft~\cite{copilot}, reliability and quality of their performance still remain controversial for healthcare applications mainly due to the data accuracy. AI in healthcare, especially related to people’s well-being and life, requires the highest accuracy, mostly higher than 80\% for diagnosis, and ranging from ~90\% to as high as 99\% in clinical data retrieval~\cite{Liu2019,Kalra2020,Rao2023,IBM_Watson_2016,WHO2021}. Critical attributes to AI’s accuracy and success are the machine learning model strength, the quantity of data inputs, and most importantly the foundation of structured high-quality data which adheres to five key criteria: completeness, correctness, concordance, plausibility, and currency~\cite{10.1136/amiajnl-2011-000681, gpai, zha2023data}.

Healthcare data composes of any information about a patient’s health and medical history, including but not limited to personal information, medical records, clinical data, and treatment plans. Clinical data, in turn, covers lab results, X-rays, diagnostic imaging, and more, of each individual, which exists in various formats, such as .txt, .doc, .xls, .pdf, DICOM, etc. In most cases, the data is stored and fragmented across multiple sources, such as devices, departments, and even at different hospitals, which poses significant challenges to doctors when accessing and gaining insights in all data for the best fit treatment decision~\cite{pirmani2024accessible, 10.1136/bmjhci-2021-100447,10.1017/cts.2022.382,10.1097/mlr.0000000000001982,10.1136/bmj.o1799}. AI technologies, which enable precise treatment and data assessment, are highly demanded to save life, especially from life-threatening and/or chronic diseases, like cancers, and COPD. Furthermore, AI technologies, with a real-time and wide integration of all individuals’ healthcare data at functional department, hospital, even region, and nation levels, will strengthen disease prediction, intervention efficiency, and cost-effectiveness.

Despite all potentials and excitements, the applications of AI in healthcare are still in their early phases, mostly small scales and trials, particularly due to legal, ethical, and regulatory considerations on data security and privacy. Key regulatory principles include explicit consent, data minimization, access and rectification rights, strong security, and breach notification. Unauthorized sharing, commercial use, and manipulation are prohibited, with strict cross-border transfer controls. Thus, processing sensitive medical information requires explicit user consent and strict compliance with laws governing data sharing between entities, in each country, such as HIPAA in the USA\cite{hipaa1996}, and GDPR in the EEA~\cite{regulation2016regulation}. 

The existing technologies-EHRs, and EMRs (Electronic Health Records, and Electronic Medical Records)-have been deployed in clinical practice, enabling clinical data collection, management, and sharing. Facile access to clinical data in these applications sparks increased interest in leveraging data for AI development. However, employing data from EHRs and EMRs in AI training has been hindered by various factors, mostly concerning their data quality and suitability. Burnum~\cite{burnum1989misinformation} reported that the adoption of EHRs has not enhanced data quality but instead resulted in a huge accumulation of poor-quality data. Overall, hospitals generate approximately 50 petabytes of data annually, 97\% of which however remains unused due to its inaccessibility and unstructured nature, leading to siloed data~\cite{weforum}. Thus, valuable insights remain trapped within individual systems, further limiting the ability to make informed decisions based on a comprehensive view of patient health.

Major industry players, such as AWS HealthLake~\cite{aws_healthlake} and Google Vertex AI~\cite{vertex_ai}, offer platforms for data processing and AI development, partly addressing the lacks in EMRs and EHRs. To our knowledge, HealthLake focuses on structured clinical data like FHIR, while Vertex AI supports general data, both structured and unstructured, yet is not specifically tailored for healthcare. Notably, the employment of these platforms could raise regulatory compliance concerns when processing sensitive data with cloud-based AI models, posing significant risks of data breaches, unauthorized access, and limited control for clients/government. Although legal clarity is still await at the time of this report, a GDPR compliance inquiry of Ireland’s Data Protection Commission into Google in 2024, regarding the development of PaLM 2, is an outstanding example~\cite{dataprotection.google}. 

To address these challenges, we have innovated AI-powered platform LizAI XT (as shown in \Cref{fig:fig1}), which is engineered with natural language processing (NLP), image processing, large language models (LLM), and advanced retrieval and data insights. LizAI XT is capable of mega-structuring a largely fragmented clinical database into a comprehensive table of all anonymized patients, and variables per disease. The clinical data mega-structure can be designed in any forms beyond table, such as charts, graphs, knowledge interaction, and more, which further provides a maximum accuracy for clinical data semantic search and management across various sources. Our technology supports both on-premises and cloud-based servers, which enables fully clients/government-controlled healthcare data, and qualifies diverse security and infrastructure needs by adhering to the strict data security, privacy, and regulatory standards, such as such as HIPAA and GDPR, of each region/country.

In this report, we design the evaluation of LizAI XT with 16 diseases, categorizing in seven groups – Oncology, Respiratory conditions, Immunological \& Allergic disorders, Neurological \& Psychiatric conditions, Infectious \& Inflammatory diseases, Reproductive Health, and Endocrine \& Metabolic disorders. Our clinically relevant database consists of a total 16,000 patients, $\sim115,000$ medical files, and $\sim800$ clinical variables, which is prepared based on real-life clinical events, guided by experts’ inputs and real-world statistics from sources, such as the CDC and NIH. The overall accuracy for data structure in LizAI XT is $>$95.0\% across all diseases, and accuracy values are exceptionally outstanding in complicated diseases, namely colorectal cancer (99.12\% $\pm$ 0.049\%), prostate cancer (99.03\% $\pm$ 0.08\%), COPD (98.89\% $\pm$ 0.076\%), and asthma (98.12\% $\pm$ 0.172), without model-overfitting. The data retrieval speed for a variable per patient is sub-second with a minimal cluster of 4x NVIDIA A30 GPU 24GB, which can exponentially be improved on more powerful GPUs. 

\input{figs/fig1.tex}

These critical attributes ensure LizAI XT potentials for launching in majority of clinics with various levels of IT infrastructure and the nationwide scalability expectation. With the addition of user-friendly interface, LizAI XT will complement the existing platforms, such as EMR/EHR, AWS HealthLake, and Google Vertex AI, for healthcare data management and AI development.

\section{Materials and Methods}
\subsection{Database Preparation}
To create highly realistic patient data for evaluating LizAI XT, we leverage the Synthea™ Patient Generator~\cite{walonoski2018synthea}, a well-established framework for producing synthetic healthcare data. Synthea’s Generic Module Framework (GMF) enables the simulation of diverse diseases and conditions, generating complete medical histories for synthetic patients from birth to the present. Each module replicates real-world clinical events, incorporating expert knowledge and statistical data from sources like the CDC~\cite{CDC2025} and NIH~\cite{NIH2025}. Additionally, the generated database adheres to standardized coding systems for laboratory results, clinical diagnoses, and medications, including LOINC~\cite{LOINC2025}, SNOMED-CT~\cite{SNOMED2025}, RxNorm~\cite{RxNorm2025}, and ICD-10~\cite{ICD102025}.

As this paper focuses on evaluating XT’s capability to structure unstructured medical data from various types and formats into datasets, we enhance Synthea-generated data by developing a component that processes its output (e.g., JSON, FHIR) using a Large Language Model (e.g., Qwen2.5~\cite{qwen2.5}, deployed on our on-premises server) to enrich the clinical context. The enriched data is then transformed into multiple widely used healthcare formats, including FHIR, HL7, CSV, PDF, TXT, as well as unstructured formats such as free-text clinical notes and imaging reports. For example, \Cref{fig:fig2} shows an example of our generated medical report for a patient.

The number of files and variables is randomized to improve data generalization and ensure a diverse representation of clinical scenarios. Overall, we created a clinically relevant database containing records from 16,000 patients across 16 diseases. In total, the database comprises 112,711 medical files in various formats, including FHIR, HL7, .csv, .pdf, and .txt, as well as unstructured formats such as free-text clinical notes and imaging reports. It includes 781 clinical variables, with definitions provided in Supplementary Information – Data S1. The diversity of data formats contributes to a comprehensive patient representation, which is essential for assessing LizAI XT’s performance.

\input{figs/fig2.tex}

\subsection{Clinical Data Mega-Structure by LizAI XT}
LizAI XT is designed to automate the structure of medical data from multiple healthcare systems, handling various data types and formats, as shown in \Cref{fig:fig3}. In particular, patient data is fragmented across multiple sources, including medical devices, departments, and even different hospitals~\cite{pirmani2024accessible, 10.1136/bmjhci-2021-100447,10.1017/cts.2022.382,10.1097/mlr.0000000000001982,10.1136/bmj.o1799}. This fragmentation poses significant challenges for doctors in accessing and synthesizing all relevant information to make the best treatment decisions. Once integrated into the healthcare infrastructure, LizAI XT automatically collects these fragmented data, structuring them into a clinical mega-structured datasets.

First, the system performs personal data anonymization (1), ensuring compliance with country/region data protection regulations, such as GDPR~\cite{regulation2016regulation}, HIPAA~\cite{hipaa1996}, and institutional policies. This process de-identifies sensitive patient information, reducing the risk of unauthorized access or data breaches. In hospital settings, where data can be shared across departments and facilities, anonymization enables secure access to lab results, imaging scans, and treatment histories while maintaining confidentiality. For example, it allows seamless collaboration between radiology, cardiology, and oncology without exposing personal details. This feature is optional and configurable, allowing institutions to tailor privacy measures to their specific regulatory and operational needs.

Then, the anonymized data is stored in a high-performance object storage system (2), which enables scalability, durability, and ability to manage diverse medical data types like imaging files, EHR records, and clinical notes. Object storage supports efficient indexing, metadata tagging, and integration with AI models and interoperability standards (FHIR, HL7). Depending on the institution’s IT and security requirements, it can be deployed on-premises or in the cloud for greater scalability, accessibility, and full control.

Following secure storage, the system intelligently routes the data to specialized processing components based on its format and type (3). These components include natural language processing (NLP) for structured and unstructured text-based data, such as physician notes, discharge summaries, and pathology reports; computer vision for analyzing medical imaging, including X-rays, MRIs, and CT scans; speech processing for transcribing and interpreting audio-based clinical records, such as doctor-patient interactions and dictated reports; and multimodal processing for integrating complex, multi-source medical data streams (4). For example, a radiology report containing both free-text descriptions and associated DICOM images can be processed using a combination of NLP and computer vision to extract clinical insights. Additionally, LizAI XT ensures interoperability by supporting standardized healthcare data formats like HL7, FHIR, and DICOM, enabling seamless integration across hospital information systems (HIS), electronic health records (EHRs), and picture archiving and communication systems (PACS). This structured routing enhances the accuracy of AI-driven analytics, ensuring that each data type is processed optimally for downstream clinical applications.

To enhance the accuracy and contextual relevance of structuring clinical variables, LizAI XT incorporates ontology-based frameworks and knowledge graphs (5), enriching its understanding of medical terminologies, relationships between clinical entities, and disease-specific variations. By leveraging standardized ontologies like LOINC~\cite{LOINC2025}, SNOMED-CT~\cite{SNOMED2025}, RxNorm~\cite{RxNorm2025}, and ICD-10~\cite{ICD102025}, the system ensures interoperability across EHRs and clinical databases while improving data consistency. This approach mitigates AI hallucinations by constraining outputs within validated medical knowledge, reducing errors and misinterpretations. For example, in oncology, it differentiates between similar terms like ``neoplasm'' and ``benign lesion'' ensuring precise clinical insights. Additionally, LizAI’s knowledge graphs enable inferential reasoning, helping identify related conditions, drug interactions, and disease progression patterns. This enhances clinical accuracy, supports decision-making, and ensures standardized data representation across diverse healthcare environments.

\input{figs/fig3.tex}

Once the data has undergone its processing pipeline, it is further refined and transformed through an advanced embedding model (6), which converts complex medical information into structured representations optimized for downstream analytics and predictive modeling. This process enhances pattern recognition, enabling more accurate disease classification, risk stratification, and treatment response predictions. The refined data is then systematically stored in a structured format, ensuring efficient retrieval for clinical interpretation, decision support, and integration with AI-driven applications, such as automated diagnostics and personalized treatment recommendations.

In the final stage, LizAI XT automatically extracts disease-relevant clinical variables (7) and structures them into comprehensive, condition-specific datasets (8), ensuring standardized and interpretable data for clinical use. These curated datasets enable precise and efficient assessments by providing a consolidated view of patient health, supporting differential diagnosis, treatment planning, and outcome prediction. Additionally, they facilitate large-scale medical research, improve predictive analytics by enhancing AI model training with high-quality inputs, and integrate seamlessly with decision support systems for real-time clinical guidance. By transforming fragmented data into structured, actionable insights, LizAI XT enhances healthcare intelligence, optimizes operational workflows, and strengthens evidence-based medical decision-making across diverse healthcare settings.

\input{figs/tab1.tex}
\input{figs/tab2.tex}

\subsection{Accuracy Assessment of Structuring Clinical Variables}
The performance of LizAI XT is assessed primarily based on the accuracy of the data structuring process for all patients across diseases. In this study, we use exact match accuracy to evaluate the correctness, which measures the proportion of fully aligned structured data with the ground truth. This metric is crucial for tasks requiring strict precision, such as clinical or regulatory data structuring, in which minor errors can impact outcomes~\cite{Tabei2008,Kim2015,Shatsky2005,Kehr2014,Chen2013,Menke2008}. The overall accuracy of LizAI XT's performance across all 16 diseases is accessed as indicated in formula (1):

\[
\mathrm{Overall\ LizAI\ XT\ Accuracy} = \frac{\sum_{i=1}^{16} \mathrm{Accuracy}_{\text{disease } i}}{16} \times 100\ (1)
\]

whereas \(\sum_{i=1}^{16} \mathrm{Accuracy}_{\text{disease }\ i}\) is the sum of accuracy values for each individual disease, which is assessed as indicated in formula (2):

\[
\mathrm{Accuracy}_{\mathrm{disease}}=\frac{\sum_{i=1}^{N}{\mathrm{Accuracy}}_{variable\ i}}{N}\times100\ (2)
\]

whereas N is the total number of clinical variables per disease, and \(\sum_{i=1}^{N} \mathrm{Accuracy}_{\text{variable}\ i}\) is the sum of accuracy values for each clinical variable of that disease. Specifically, for each disease, we calculate the accuracy for each variable using the following formula (3):

\[
\mathrm{Accuracy}_{\mathrm{variable}}=\frac{\sum_{i=1}^{M}1\left(\widehat{X_i}=X_i\right)}{M}\times100\ (3)
\]

whereas \(X_i\) is the ground truth value for clinical variable \(i\), \(\widehat{X_i}\) is the extracted value by LizAI XT for clinical variable \(i\), \(1\left(\widehat{X_i} = X_i\right)\) equals 1 if correctly extracted and 0 otherwise, and \(N\) is the total number of patients.

\paragraph{Standard Error of Accuracy} We assess the reliability of the accuracy measurement with standard error (SE) of accuracy, which represents the possible variation in accuracy across different samples. The standard error is calculated as indicated in formula (4):

\[
SE=\sqrt{\frac{p\left(1-p\right)}{N}}\ (4)
\]

whereas p is the accuracy proportion (e.g., 0.95 for 95\% accuracy), N is the total number of samples, SE quantifies the uncertainty in the accuracy estimate. For instance, if LizAI XT achieves 95\% accuracy (0.95) across 1,000 patients, the standard error is 0.69\%. This means the true accuracy is expected to fall within $\pm$ 0.69\% of the reported value, indicating a high level of confidence in the measurement.

\section{Results}
\subsection{Preparation of Clinically Relevant Database for LizAI XT Performance Evaluation}
We generated a clinically relevant database, consisting records of a total 16,000 patients, in 16 diseases. \Cref{tab:tab1} lists all studied diseases, each of which has 1,000 patients, thousands of medical files, and a plenty of clinical variables. These variables are modeled real-life clinical events as guided by experts inputs and real-world statistics from health organizations, such as CDC~\cite{CDC2025} and NIH~\cite{NIH2025}. Furthermore, this database follows standardized coding for laboratory results, clinical diagnoses, and medications, such as LOINC~\cite{LOINC2025}, SNOMED-CT\cite{SNOMED2025}, RxNorm~\cite{RxNorm2025}, and ICD-10\cite{ICD102025}.

The number of files and variables are randomized to ensure a realistic representation of clinical scenarios. Notably, the database of 112,711 medical files includes multiple types and formats, such as FHIR, HL7, .cvs, .pdf, .txt, as well as unstructured formats, like free-text clinical notes, and imaging reports. And there are 781 clinical variables whose definitions are given in Supplementary Information – file S1. The variation of data formats renders a comprehensive patient representation, which is crucial for LizAI XT performance assessments. While variables varies across diseases based on complexity, data availability, and diagnostic requirements, enabling condition-specific evaluations. 

In this study, we employ this database to evaluate different technical aspects of LizAI XT, including performance assessment in different diseases, overall accuracy, cross-checking various clinical variable groups, and identifying outliers to understand challenges that can help to improve data processing and enhance model robustness. Additionally, we categorized 781 clinical variables into ten sub-groups, as presented in \Cref{tab:tab2}, to investigate the adaptability of LizAI XT across diseases and variable types.

\subsection{Clinical Data Mega-Structure by LizAI XT – A Case of Prostate Cancer}
LizAI XT efficiently mega-structures all clinical data per disease into datasets by relevant variables (as reported in \Cref{tab:tab1}, and Supplementary Information – file S1), from the vastly fragmented database of 16,000 patients, ~115,000 medical files, and ~800 clinical variables. \Cref{fig:fig4} illustrates the data processing in LizAI XT and data-tables, as outputs, of some representative diseases. Our platform also includes the fully anonymized procedure, which guarantees the data privacy policies. Additionally, it is important to note that the outputs can be designed in any forms, such as graphs, and knowledge relationship.   
 
We additionally present a simplified data-table example of all prostate cancer patients mega-structured by some selected variables in \Cref{tab:tab3}.

\input{figs/fig4.tex}

\input{figs/tab3.tex}

\section{LizAI XT Performance Evaluation }

\subsection{Overall LizAI XT Performance Accuracy}
In our latest assessment, LizAI XT achieves an overall accuracy of 95.79\% $\pm$ 5.69\% when structuring a database of 16,000 patients, $\sim$115,000 medical files, and 781 clinical variables, into datasets per disease by relevant variables (\Cref{fig:fig5}). This result demonstrates the platform’s effectiveness across multiple diseases and fragmented database. Notably, LizAI XT performance reaches the highest accuracy in complicated cases of colorectal cancer (99.12\% $\pm$ 0.049\%), prostate cancer (99.03\% $\pm$ 0.08\%), COPD (98.89\% $\pm$ 0.076\%), and contraceptives (98.28\% $\pm$ 0.12\%), each of which holds complexity of clinical databases and various numbers of variables. Only ear infections and bronchitis show sub-ninety accuracy, at 87.92\% $\pm$ 0.176\% and 87.57\% $\pm$ 0.431\%, respectively, which might be caused by the outliers with accuracy below 85\% due to the variability in clinical variables, such as broad symptomatology, and overlapping diagnostic criteria (lists of variables in each disease are provided in Supplementary Information – file S1). Some examples of broad symptomatology include variables in bronchitis symptoms, such as cough and shortness of breath, which may overlap with other respiratory conditions like asthma or pneumonia. While there are some inconsistencies in ear infections records, as they may be documented as otitis media, ear pain, or effusion. 

\input{figs/fig5.tex}

\subsection{Analysis of Outliers in Accuracy and Their Impacts on the Overall LizAI XT Performance}
We selected accuracy below 85\% as outliers based on SE, which impact the performance of LizAI XT in some diseases, such as bronchitis, and ear infections. Notably, all variables in colorectal cancer, prostate cancer, ADD, and COPD population were structured at accuracy higher than 85\% and thus not included in the latter analysis. \Cref{fig:fig6}A reports the total outliers and their proportion in each disease, with majority below 10\% (lists of outliers in each disease are provided in Supplementary Information – file S2). Despite their minor presentation, as the total 45 outliers across 16 diseases only account for 5.76\% in nearly 800 variables (\Cref{fig:fig6}B), we further investigate the impact of these outliers in order to improve the platform’s performance. We thus deep-dive into these 45 outliers by categories and identify the variables which contribute the most to the outlier group. Interestingly, names (medical), conditions, observations, care-plans, and device variables are not among the outliers, indicating highly accurate performance in these categories (\Cref{fig:fig6}C). The three largest contributors among outliers are symptoms, medications, and immunizations at 40\%, 26.47\%, and 24.44\%, respectively. 

We further examine their average accuracy of outlier variable categories to assess potential impacts and quantify their extent. While these outliers are lower compared to the maximum accuracy of 99\%, their performance remains within an acceptable range ($\sim$63\% to $\sim$85\%), as shown in \Cref{fig:fig6}D, indicating that LizAI XT’s AI-powered model maintains a reasonable level of reliability. Notably, although the accuracy of outliers is statistically significant when compared to the group of variables with accuracy above 85\%. However statistical tests such as the t-test~\cite{student1908probable}, Mann-Whitney U test~\cite{mann1947test}, and bootstrapping~\cite{efron1979bootstrap} indicate a significant difference between outliers (scores below 85\%) and the rest of the data, the practical impact is minimal. Cohen’s d of -0.26 falls within the small effect range ($|d| < 0.3$), meaning the shift in mean accuracy is minor and does not substantially affect LizAI XT’s overall performance~\cite{cohen1988statistical}, as shown in \Cref{fig:fig7}. \Cref{fig:fig6}E in turn illustrates impact of the outliers on different diseases. Immunization and symptom categories appear as outliers in a larger number of diseases, while the other only affect a few.

\input{figs/fig7.tex}
\input{figs/fig6.tex}

\subsection{Speed of LizAI XT in Data Mega-Structure}
We note that the investigation was performed on a 4x NVIDIA A30 (24GB) GPU setup. In our latest assessment, LizAI XT efficiently achieves sub-second processing speed even with the minimal cluster power. The data retrieve speed is inference time per clinical variable per patient, and we continuously track GPU utilization using NVIDIA-SMI and measure comparison time to evaluate how quickly LizAI XT structures and matches clinical variables to the ground truth. The data clearly demonstrates the platform’s capability for real-time clinical data structure.

\section{Discussion and Conclusion}
Healthcare data fragmentation at all levels, from functional departments, clinics/hospitals, health organizations like CDC, to government, poses a significant challenge for AI-driven healthcare innovations, as machine learning models require structured and standardized data for meaningful insights~\cite{pirmani2024accessible, 10.1136/bmjhci-2021-100447,10.1017/cts.2022.382,10.1097/mlr.0000000000001982,10.1136/bmj.o1799,Reddy2021,Gilvaz2019,Khan2023,Ogaga2023}. AI technologies, which enable real-time, accurate, and efficient data integration of all individual’s healthcare data, are urgently needed. In this report, we introduce an innovative platform LizAI XT which mega-structures all fragmented databases from different sources into datasets and unlock clinically relevant information, thereby enabling advanced analytics, clinical decision support, and precision medicine. Furthermore, structured datasets per disease can accelerate scientific and technological innovations, improves diagnostic accuracy, reduces physician burnout, enhances precision treatment, and optimizes management across all levels of healthcare, from clinics and insurance companies to government.

LizAI XT’s clinical data mega-structure was accessed on a fragmented database of 16,000 patients, $\sim$800 clinical variables, and 115,000 medical files in different types and formats. Overall, LizAI XT is a robust and reliable platform consistently achieving average 99\% accuracy, especially in complicated and chronic diseases, including colorectal cancer, prostate cancer, and COPD. Importantly, the database prepared for the LizAI XT performance assessment was completely blinded and has never been exposed to the platform’s AI-powered model. Thus, the accuracy values truly reflects LizAI XT applicability in diseases beyond the studied list without relying on an established memorization, confirming our data reliability for real-world clinical applications without overfitting. Sub-ninety accuracy values are recorded in some diseases, such as ear infections and bronchitis, due to the contribution of outlier variables – below 85\%. The performance in these outlier variables is likely impacted by the technical challenges, such as ambiguity, overlap, inconsistent formatting, and broad symptomatology. Medical codes like "Encounter for problem", for example, are broad and can apply to multiple conditions, making precise classification challenging. Similarly, names such as "Encounter Module Scheduled Wellness" may contain multiple components that vary across documentation systems, leading to inconsistencies. Additionally, some terms have multiple meanings depending on the context, and variations in healthcare documentation further complicate standardization. These factors make it harder for the model to distinguish between similar entities, requiring improved context-aware processing, entity linking, and standardization techniques to enhance accuracy. Improving specificity and standardization in these areas could further enhance LizAI XT's performance. 

LizAI XT is optimized for speed and real-time processing on a minimal server setup, utilizing 4x A30 24GB GPUs. Even in this streamlined environment, our platform can process at sub-second speed per variable per patient. On the more powerful infrastructures, such as dual Intel Xeon Platinum processors, 1TB RAM, and NVMe SSD storage, LizAI XT scales seamlessly for hospital networks, research institutions, and national healthcare databases. Data structure speed can increase by three times with A100 GPUs, while upgrading to H100 GPUs can further enhance the performance by six times, enabling near-instantaneous large-scale medical data processing. Together with the demonstrated high accuracy, LizAI XT claims its capability for launching in various IP infrastructure and nationwide scalability.
The increasing digitalization in healthcare has led to widespread adoption of EHRs/EMRs worldwide, yet a significant portion of clinical data remains fragmented, unstructured, and underutilized. AI-powered data mega-structure by LizAI XT complements these EMRs/EHRs systems by addressing their limitations in data processing, especially the lacks of deep AI integration for contextualizing and structuring free-text medical records ~\cite{ramakrishnaiah2024ehr, alanazi2023clinicians, deliberato2017clinical, ronquillo2022untapped}. Clinical data of cancer patients as an example can be stored in different hospitals’ EHRs and originated from different sources. Important details, such as tumor diagnosis and progression, prior treatments, and doctors' notes, are trapped in various file types and formats, some of which are healthcare system notes, .pdf, handwritten notes, DICOM, and .xls files. Oncologists need all data insights for a treatment plan but spend a significant amount of time manually searching and reading, possibly in EHRs. This reality negatively impacts timely decision-making, despite having EHRs. Employing LizAI XT enables the structure of a single consolidated data-table, or any another designed formats, which includes all meaningful data of not only one but all patients. In this case, oncologists can efficiently make the best fit treatment decision in a timely manner, instead of going through thousands of medical files with risk of missing important information. 

Importantly, LizAI XT enables fully clients/government-controlled healthcare data, which can be installed both on-premises and cloud-based servers, and qualifies diverse security and infrastructure needs by adhering to the strict data security, privacy, and regulatory standards, of each region/country. In addition to the clinical specificity, LizAI XT’s features could outperform the conventional AI for healthcare applications and the current approach by big players, such as Amazon, and Google. And our technology can fulfill the needs of AI developments.

\section*{Acknowledgments}

\section*{Funding}
This research received no external funding.

\section*{Author contributions} 
Trung Tin Nguyen and David R. Elmaleh contributed equally to this article and in all the following categories: conceptualization; methodology; software; validation; formal analysis; investigation; resources; data curation; writing—original draft preparation; writing—review and editing; visualization; supervision; project administration; and funding acquisition. Prof. MD. Salomon M. Stemmer contributed to this article and in the following categories: writing—original draft preparation; writing—review and editing; additionally, Prof. MD. Stemmer involved in the data outcome’s design which reflects the institutions’ and physicians’ needs in healthcare assets management.  All authors have read and agreed to the published version of the manuscript.

\section*{Competing interests} 
Authors Trung Tin Nguyen and David R. Elmaleh were co-founders of LizAI Inc. Authors Trung Tin Nguyen and David R. Elmaleh are inventors and have filed the US patent application number 19/087,980, which claims intellectual property for the platform published herein.

\section*{Data and materials availability}
All data are included in the manuscript and Supplementary Materials. Additional data and information are available upon request.

\section*{References}

\bibliography{mybibfile}


\end{document}

%% file: figs/fig1.tex
\begin{figure}[t]
    \centering
    \includegraphics[width=1\linewidth]{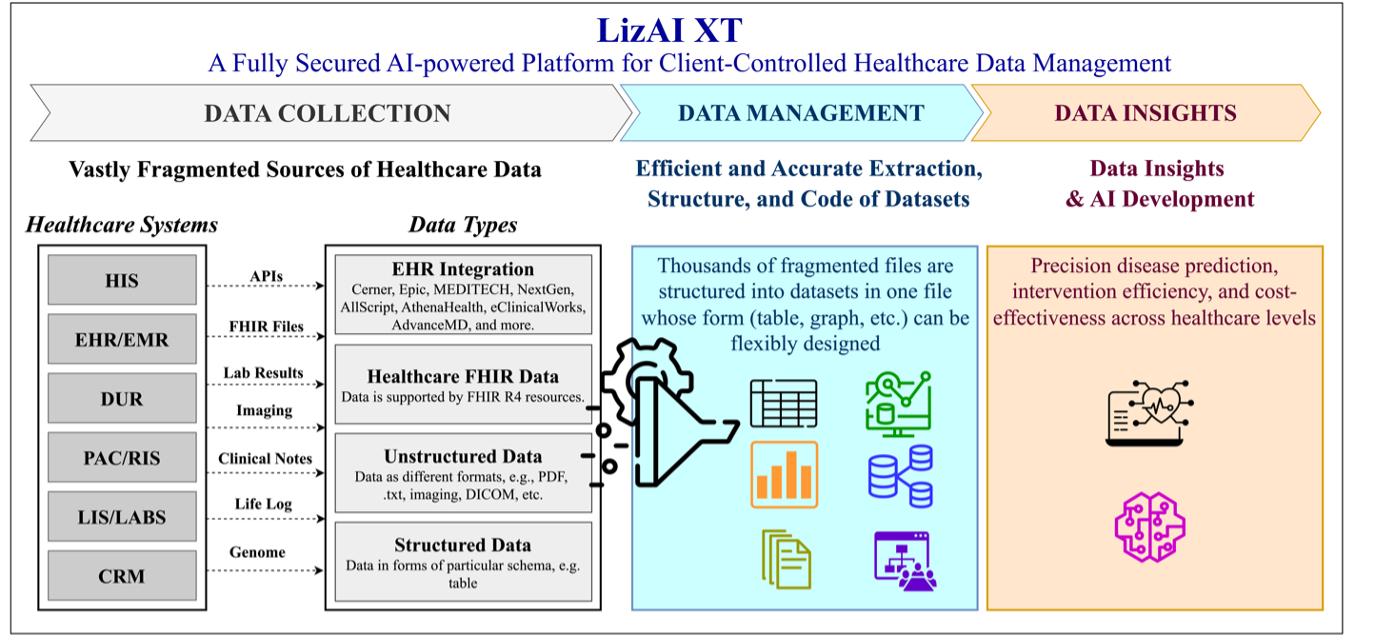}
    \caption{Overall illustration of a fully secured AI-powered LizAI XT Platform for client-controlled healthcare data management. This report focuses on the data management by clinical data mega-structure using LizAI XT.}
    \label{fig:fig1}
\end{figure}

%% file: figs/fig2.tex
\begin{figure}[H]
    \centering
    \includegraphics[width=1\linewidth]{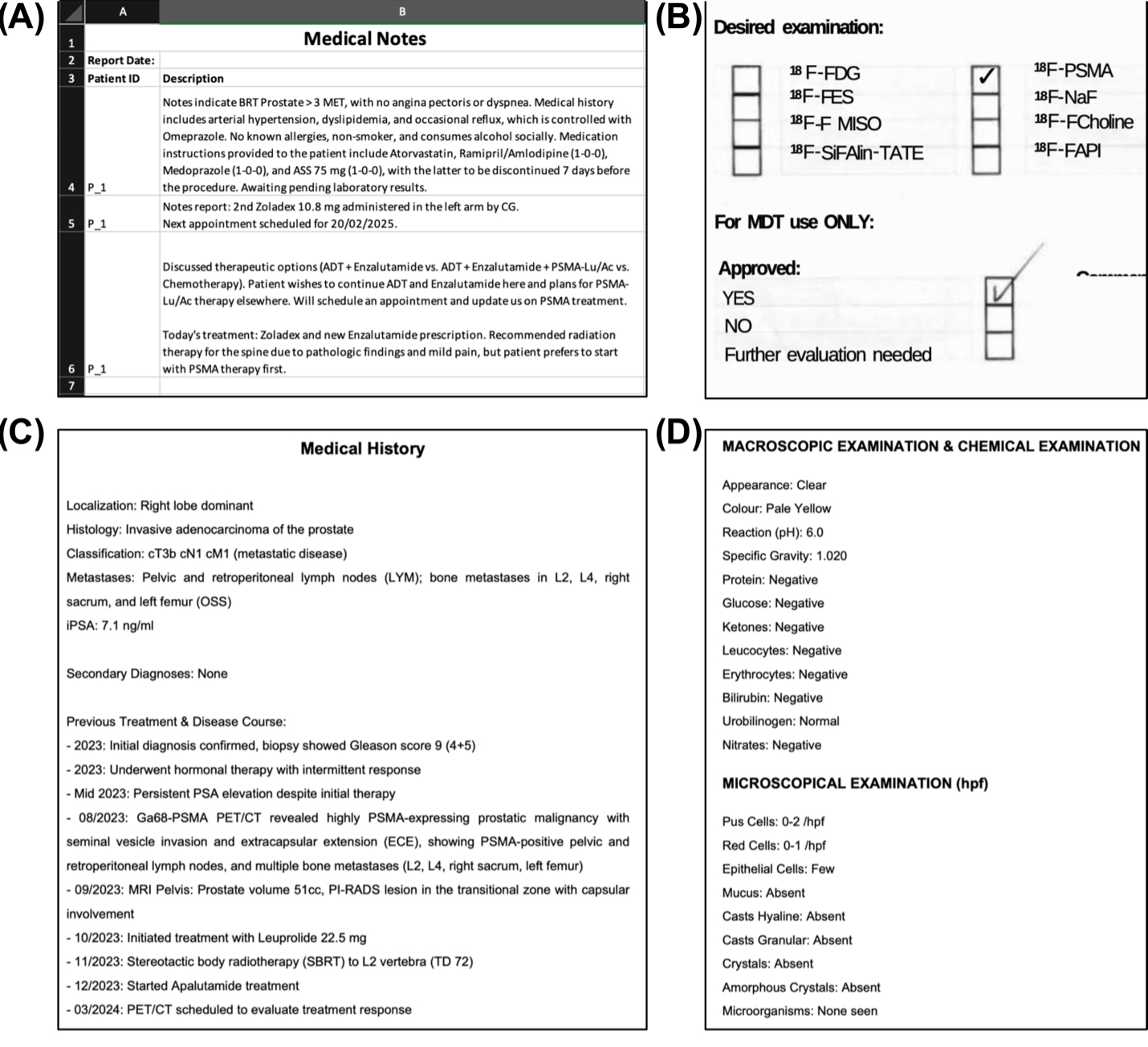}
    \caption{Some examples of our generated medical files for a patient: (A) a spreadsheet containing key clinical notes from doctors, (B) a scanned PDF of a checklist, (C) a printed PDF report summarizing the patient's complete medical history, and (D) a PDF report of laboratory test results, covering macroscopic, microscopic, and chemical examinations.}
    \label{fig:fig2}
\end{figure}

%% file: figs/fig3.tex
\begin{figure}[t]
    \centering
    \includegraphics[width=1\linewidth]{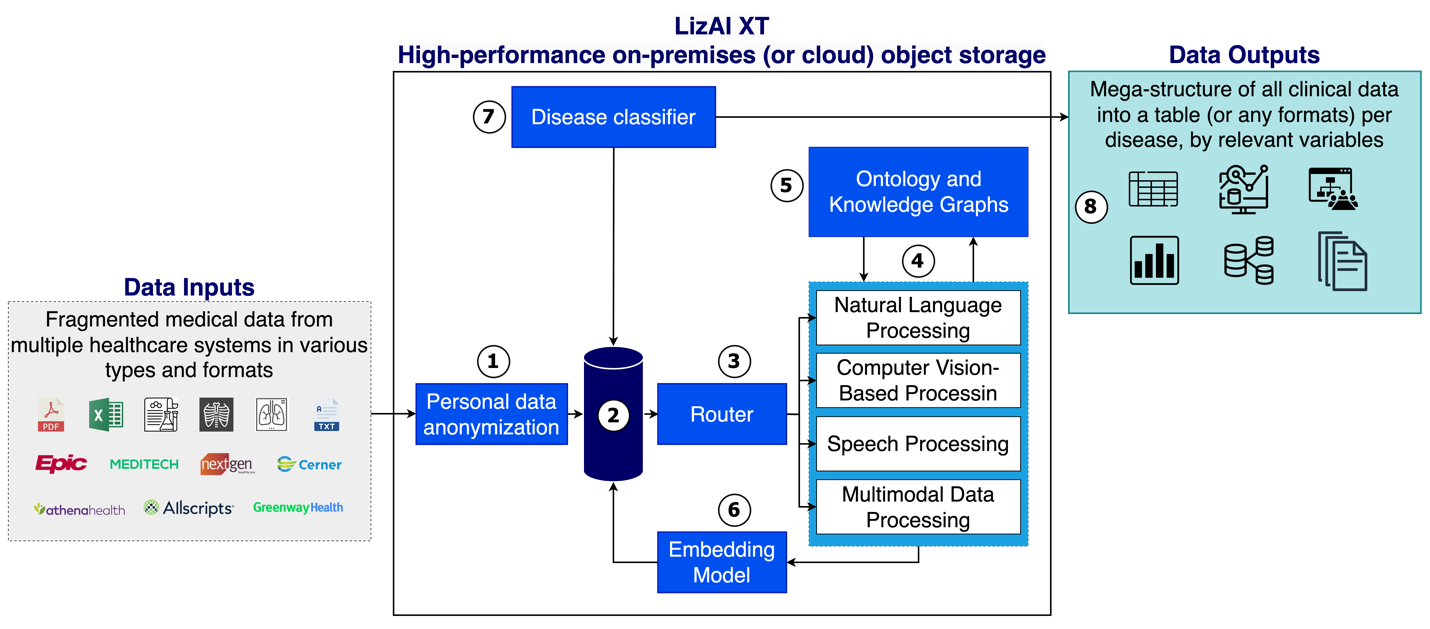}
    \caption{System diagram of Clinical Data Mega-Structure by LizAI XT which automates the data collection, anonymization, storage, and structuring of medical data from various healthcare systems. The system de-identifies personal data \textbf{(1)}, stores it in a high-performance object storage system \textbf{(2)}, and routes it to specialized processing components based on data type \textbf{(3)}. NLP, computer vision, speech processing, and multimodal analysis \textbf{(4)} are enhanced by ontology and knowledge graphs \textbf{(5)} for improved accuracy. Processed data undergoes refinement \textbf{(6)} before clinical variables are extracted \textbf{(7)} and structured into disease-specific datasets for research, analytics, and decision support \textbf{(8)}.}
    \label{fig:fig3}
\end{figure}

%% file: figs/tab1.tex
\begin{table}[H]
\footnotesize
\centering
\caption{Summaries of diseases, as well as number of patients, number of medical files, and number of clinical variables for each disease in our database. This clinically relevant database is a random mix of various clinical data types and formats, such as FHIR, HL7, .cvs, .pdf, .txt, free-text clinical notes, and imaging reports, which is generated as guided by experts inputs and real-world statistics from health organizations, such as CDC and NIH. The database is used for LizAI XT performance evaluation in data mega-structure in this report. The lists of clinical variables for each disease are given in Supplementary Information – file S1.}
\label{tab:tab1}
\begin{tabular}{@{}lp{4.4cm}ccc@{}}
\toprule
\textbf{Disease}                          & \textbf{Short Description}                                                 & \multicolumn{1}{c}{\textbf{\begin{tabular}[c]{@{}c@{}}Number \\ of\\ Patients\end{tabular}}} & \multicolumn{1}{c}{\textbf{\begin{tabular}[c]{@{}c@{}}Number of \\ Medical \\ Files\end{tabular}}} & \multicolumn{1}{c}{\textbf{\begin{tabular}[c]{@{}c@{}}Number of \\ Clinical \\ Variables\end{tabular}}} \\ \midrule
\textbf{Colorectal Cancer}                & Cancer affecting the colon or rectum.                                      & 1,000                       & 5,317                            & 105                                   \\
\textbf{Prostate Cancer}                  & A common male cancer in the prostate gland.                                & 1,000                       & 22,781                           & 50                                    \\
\textbf{Contraceptives}                   & Medications or devices used for birth control.                             & 1,000                       & 5,718                            & 67                                    \\
\textbf{Female Reproduction}              & Conditions related to women's reproductive health.                         & 1,000                       & 5,102                            & 25                                    \\
\textbf{Gout}                             & Arthritic condition caused by uric acid crystal buildup in joints.         & 1,000                       & 1,492                            & 41                                    \\
\textbf{Attention Deficit Disorder (ADD)} & Neurodevelopmental disorder affecting focus and impulse control.           & 1,000                       & 5,549                            & 41                                    \\
\textbf{Epilepsy}                         & Neurological disorder causing recurrent seizures.                          & 1,000                       & 6,279                            & 36                                    \\
\textbf{COPD}                             & Progressive lung disease causing breathing difficulties.                   & 1,000                       & 5,327                            & 75                                    \\
\textbf{Asthma}                           & Chronic condition causing airway inflammation and difficulty breathing.    & 1,000                       & 8,360                            & 67                                    \\
\textbf{Allergic Rhinitis}                & Inflammation of nasal passages due to allergens.                           & 1,000                       & 5,397                            & 42                                    \\
\textbf{Bronchitis}                       & Inflammation of bronchial tubes, leading to coughing and mucus production. & 1,000                       & 11,991                           & 51                                    \\
\textbf{Dermatitis}                       & Inflammation of the skin causing redness and itching.                      & 1,000                       & 5,229                            & 42                                    \\
\textbf{Atopy}                            & Genetic tendency to develop allergic conditions.                           & 1,000                       & 4,996                            & 25                                    \\
\textbf{Food Allergies}                   & Immune response triggered by certain foods.                                & 1,000                       & 5,480                            & 35                                    \\
\textbf{Appendicitis}                     & Inflammation of the appendix, often requiring surgery.                     & 1,000                       & 5,322                            & 43                                    \\
\textbf{Ear Infections}                   & Infections of the middle ear, causing pain and fluid buildup.              & 1,000                       & 8,371                            & 36                                    \\
\textbf{Total}                            &                                                                            & 16,000                      & 112,711                          & 781                                   \\ \bottomrule
\end{tabular}
\end{table}

%% file: figs/tab2.tex
\begin{table}[H]
\centering
\caption{The total $\sim$800 clinical variables in all 16 diseases are categorized in ten groups, which supports the adaptability assessment of LizAI XT across diseases and variable types.  }
\label{tab:tab2}
\footnotesize
\begin{tabular}{@{}llp{5cm}@{}}
\toprule
\textbf{Variables Categories}     & \textbf{Descriptions}                          & \textbf{Examples}                                                             \\ \midrule
\textbf{Immunizations}            & Administered vaccines of any kinds             & DTaP, Influenza vaccine                                                       \\
\textbf{Codes (medical)}          & Medical encounter and/or procedure identifiers & Encounter for check-up, Death Certification                                   \\
\textbf{Names (medical)}          & Titles of medical encounters                   & Chemotherapy Encounter, Routine Colonoscopy Encounter                         \\
\textbf{Medications (treatments)} & Prescribed drugs or treatments                 & Oxaliplatin Injection, Leucovorin Injection                                   \\
\textbf{Symptoms}                 & Reported health complaints                     & Abdominal Pain, Fatigue                                                       \\
\textbf{Conditions}               & Diagnosed diseases and/or disorders            & Anemia (disorder), Malignant tumor of colon                                   \\
\textbf{Observations}             & Recorded health measurements                   & Hemoglobin level, Pain severity score                                         \\
\textbf{Care-plans}               & Structured treatment or health plans           & Cancer care plan, Healthy diet                                                \\
\textbf{Procedures}               & Medical interventions or diagnostics           & Colonoscopy, Biopsy of colon                                                  \\
\textbf{Devices (methods)}        & Medical equipment for patient use.             & Oxygen concentrator (physical object), Wheelchair accessory (physical object) \\ \bottomrule
\end{tabular}
\end{table}

%% file: figs/fig4.tex
\begin{figure}[t]
    \centering
    \includegraphics[width=1\linewidth]{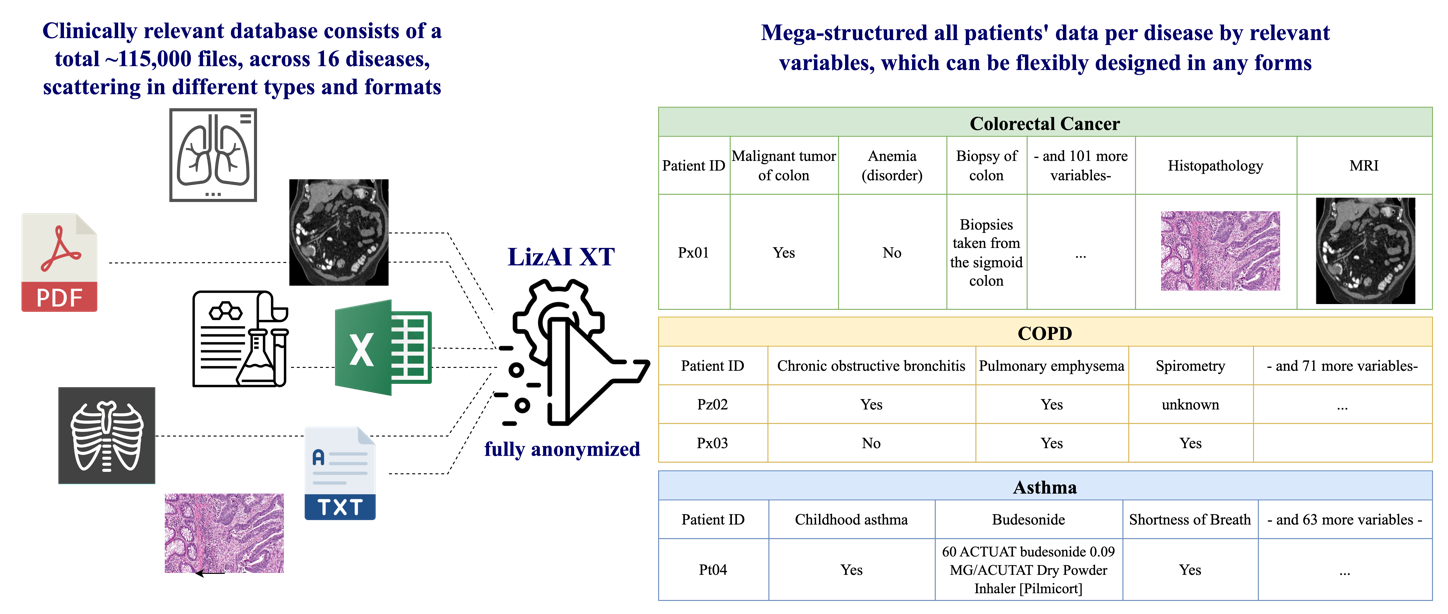}
    \caption{Illustration of data mega-structure by LizAI XT. Thousands of files and information of 16,000 patients in this study are fragmented in different types and formats, which can efficiently be structured into one data-table per disease by relevant variables. }
    \label{fig:fig4}
\end{figure}

%% file: figs/tab3.tex
\begin{landscape}
\begin{center}

\begin{table}[]
\centering
\caption{We present a simplified mega-structured data-table of all prostate cancer patients by some selected variables. This mega-structured data-table allows semantic search and insights in all clinical data from a vastly fragmented sources. This outcome can be designed in any forms, such as graphs, knowledge relationship, and more. All data can be linked to the original sources for validation, and all clinical images can be shown in one page. Abbreviations: iPSA – Initial prostate-specific antigen level; ISUP – International Society of Urological Pathology grade, ADT – Androgen Deprivation Therapy.}
\label{tab:tab3}
\resizebox{24cm}{!}{%
\begin{tabular}{@{}|l|l|l|l|l|l|l|l|l|l|@{}}
\toprule
Anonymized ID   & \textbf{iPSA} & \textbf{ISUP score in biopsy specimen} & \textbf{Date of biopsy} & \textbf{Imaging for primary staging} & \textbf{ADT duration} & \textbf{Other systemic therapy primary treatment} & \textbf{Radiation prostate} & \textbf{Number of pelvic lymph nodes in imaging} & \textbf{Type of local salvage treatment}                  \\ \midrule
\textbf{P\_844} & 44 ng/ml      & 8 (3+5)                                & 2017-07-31              & PSMA-PET/CT or PET/MR                & None                  & Enzalutamide                                      & None                        & 5                                                & Radiotherapy of the thoracic segment of the spinal column \\ \midrule
\textbf{P\_33}  & 30 ng/ml      & 10 (5+5)                               & 2015-12-11              & MRI OF THE PELVIS-PROSTATE           & None                  & None                                              & Yes                         & None                                             & HDR                                                       \\ \midrule
\textbf{P\_272} & 8.8 ng/ml     & 4                                      & 2020-05-16              & MRI OF THE PELVIS-PROSTATE           & 9 months              & Enzalutamide                                      & Yes                         & None                                             & None                                                      \\ \midrule
\textbf{P\_229} & 42 ng/ml      & 5                                      & unknown                 & PET/CT scan                          & 5 months              & Enzalutamide                                      & Yes                         & None                                             & Conventional-fractionation IMRT combined with HDR         \\ \midrule
\textbf{P\_478} & 47 ng/ml      & 7 (3+4)                                & unknown                 & PET/CT imaging                       & None                  & None                                              & Yes                         & 2                                                & SBRT plus HDR                                             \\ \midrule
\textbf{P\_32}  & 18 ng/ml      & 3                                      & 2019-09-27              & PET/CT                               & 9 months              & Enzalutamide                                      & Yes                         & None                                             & None                                                      \\ \midrule
\textbf{P\_441} & 38 ng/ml      & 3                                      & unknown                 & MRI OF THE PELVIS-PROSTATE           & None                  & None                                              & Yes                         & None                                             & SBRT plus HDR for 2 months                                \\ \midrule
\textbf{P\_221} & 8.8 ng/ml     & None                                   & 02.11.2022              & MRI OF THE PELVIS-PROSTATE           & 4 years               & None                                              & Yes                         & None                                             & Brachytherapy                                             \\ \midrule
\textbf{P\_844} & 44 ng/ml      & 8 (3+5)                                & 2017-07-31              & PSMA-PET/CT or PET/MR imaging        & None                  & Enzalutamide                                      & NONE                        & 5                                                & Radiotherapy of the thoracic segment of the spinal column \\ \midrule
\textbf{P\_33}  & 30 ng/ml      & 10 (5+5)                               & 2015-12-11              & MRI OF THE PELVIS-PROSTATE           & None                  & None                                              & Yes                         & None                                             & HDR                                                       \\ \midrule
\textbf{P\_227} & 28 ng/ml      & None                                   & 2017-02-15              & PET/CT scan                          & None                  & None                                              & Yes                         & 2                                                & IMRT (Intensity-Modulated Radiation Therapy)              \\ \midrule
\textbf{P\_424} & 41 ng/ml      & None                                   & 2002-06-22              & MRI OF THE PELVIS-PROSTATE           & None                  & Enzalutamide                                      & None                        & None                                             & STRING: Brachytherapy                                     \\ \midrule
\textbf{P\_673} & 8.8 ng/ml     & None                                   & 2003-09-09              & PSMA-PET/CT or PET/MR                & None                  & None                                              & Yes                         & None                                             & Brachytherapy monotherapy                                 \\ \midrule
\textbf{P\_272} & 8.8 ng/ml     & 4                                      & 2020-05-16              & MRI OF THE PELVIS-PROSTATE           & 9 months              & Enzalutamide                                      & Yes                         & None                                             & None                                                      \\ \midrule
\textbf{P\_229} & 42 ng/ml      & 5                                      & unknown                 & PET/CT scan                          & 5 months              & Enzalutamide                                      & Yes                         & None                                             & Conventional-fractionation IMRT combined with HDR         \\ \midrule
\textbf{P\_478} & 47 ng/ml      & 7 (3+4)                                & unknown                 & PET/CT imaging                       & None                  & None                                              & Yes                         & 2                                                & SBRT plus HDR                                             \\ \midrule
\textbf{P\_32}  & 18 ng/ml      & 3                                      & 2019-09-27              & PET/CT                               & 9 months              & Enzalutamide                                      & Yes                         & None                                             & None                                                      \\ \midrule
\textbf{P\_441} & 38 ng/ml      & 3                                      & unknown                 & MRI OF THE PELVIS-PROSTATE           & NONE                  & NONE                                              & Yes                         & None                                             & SBRT plus HDR for 2 months                                \\ \bottomrule
\end{tabular}%
}
\end{table}
\end{center}
\end{landscape}

%% file: figs/fig5.tex
\begin{figure}[H]
    \centering
    \includegraphics[width=0.8\linewidth]{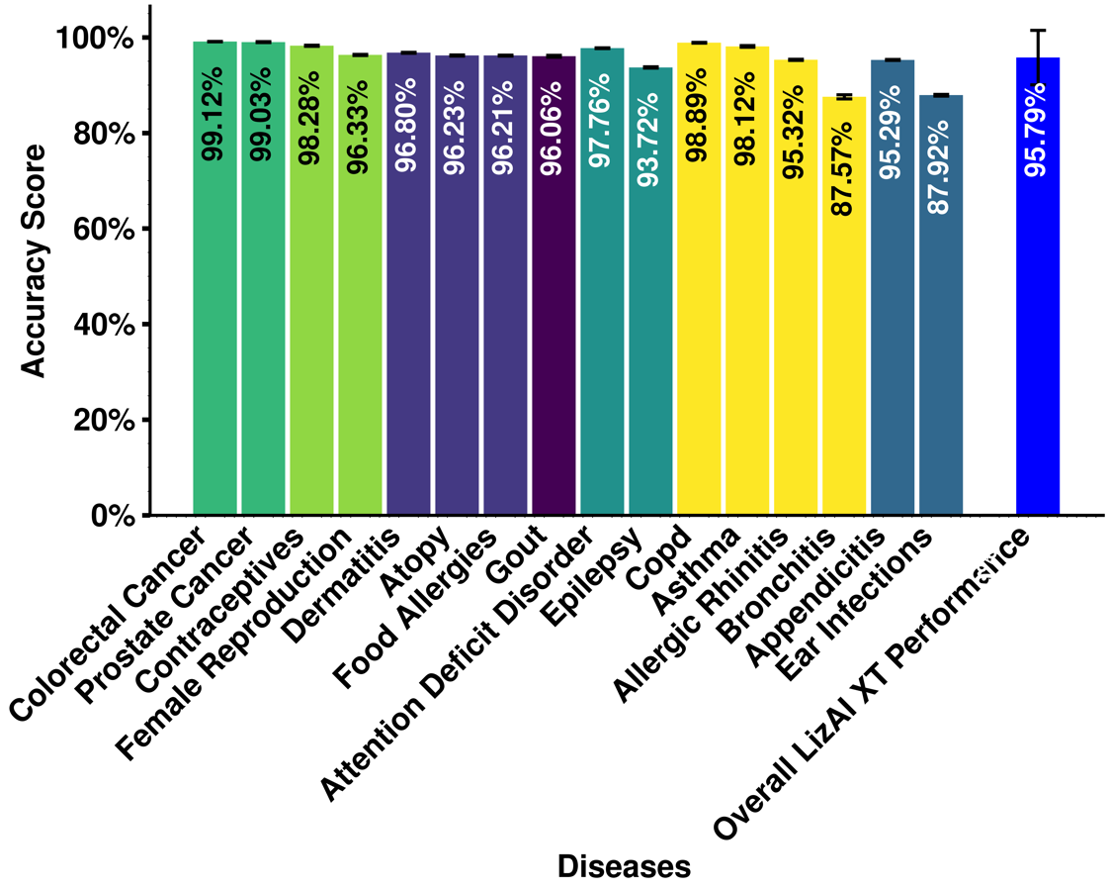}
    \caption{Assessment of LizAI XT’s performance based on accuracy of data mega-structure. The accuracy is calculated for each disease, and the overall performance is calculated for structuring the entire database.}
    \label{fig:fig5}
\end{figure}

%% file: figs/fig7.tex
\begin{figure}[H]
    \centering
    \includegraphics[width=0.8\linewidth]{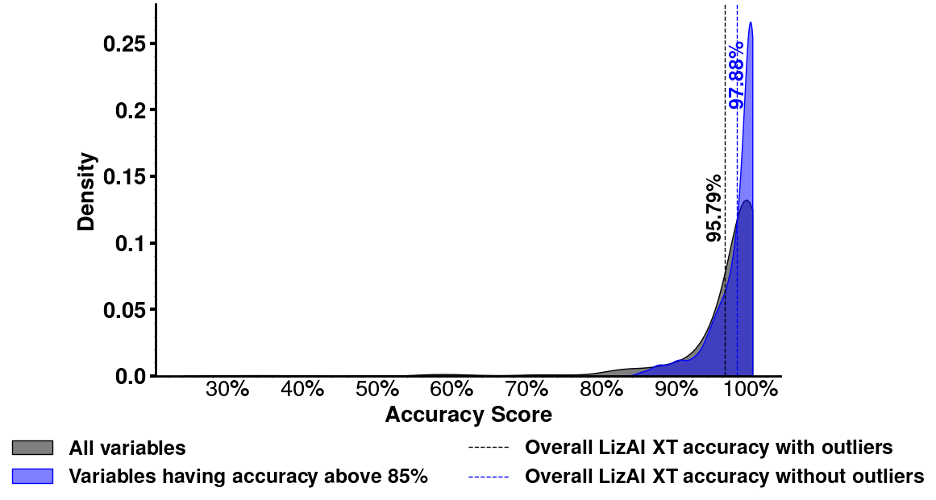}
    \caption{Visualizing the distributions of mean scores before and after removing outliers shows overlapping curves, reinforcing that their exclusion does not meaningfully alter overall accuracy. A truly significant impact would require Cohen's d $>$ 0.5 (medium) or 0.8 (large), which is not observed here.}
    \label{fig:fig7}
\end{figure}

%% file: figs/fig6.tex
\begin{figure}[H]
    \centering
    \includegraphics[width=0.99\linewidth]{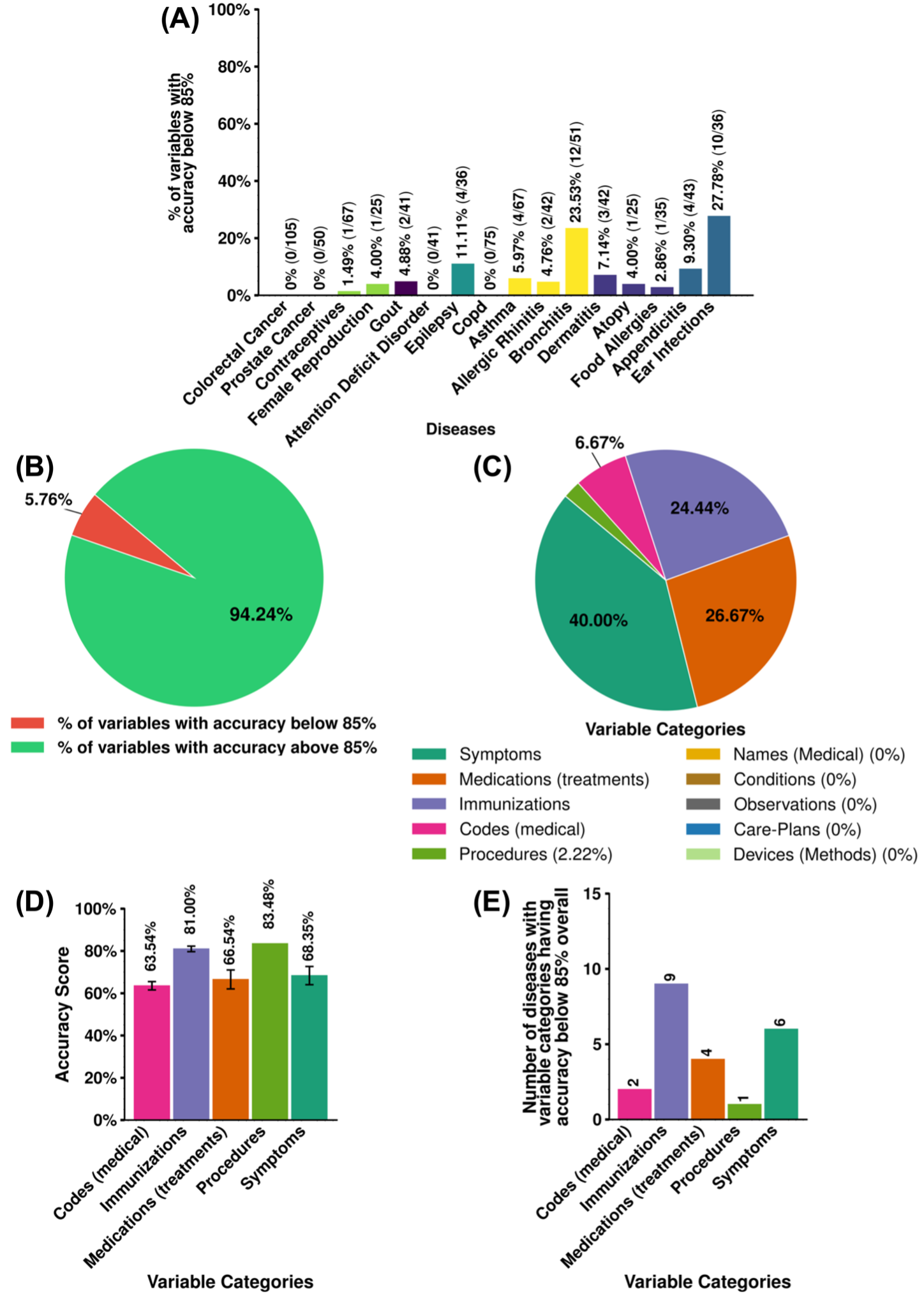}
    \caption{Assessment of 45 outliers (accuracy below 85\%) in LizAI XT's performance across 16 diseases. (A) Numbers and portions (in percentage) of outlier variables in each disease. Among 45 outlier variables, we calculate (B) the presentation of 45 outliers variables versus a total 781 clinical variables in all diseases; (C) the contribution of each variable categories (in percentage); and (D) the average accuracy score of each variable categories. (E) Impact of each outlier category on group of 16 diseases.}
    \label{fig:fig6}
\end{figure}